# Monopole and Topological Electron Dynamics in Adiabatic Spintronic and Graphene Systems

S. G. Tan,[1] M. B. A. Jalil,[2] Takashi Fujita[1,2]

[1]Data Storage Institute, Agency for Science, Technology and Research (A*STAR)
DSI Building, 5 Engineering Drive 1, Singapore 117608

[2]Information Storage Materials Laboratory, Electrical and Computer Engineering Department, National University of Singapore, 4 Engineering Drive 3, Singapore 117576

Abstracts

A unified theoretical treatment is presented to describe the physics of electron dynamics in semiconductor and graphene systems. Electron spin's fast alignment with the Zeeman magnetic field (physical or effective) is treated as a form of adiabatic spin evolution which necessarily generates a monopole in magnetic space. One could transform this monopole into the physical and intuitive topological magnetic fields in the useful momentum (K) or real spaces (R). The physics of electron dynamics related to spin Hall, torque, oscillations and other technologically useful spinor effects can be inferred from the topological magnetic fields in spintronic, graphene and other SU(2) systems.

Contact:
Dr S. G. Tan
Division of Spintronic, Media and Interface
Data Storage Institute
(Agency for Science and Technology Research)
DSI Building 5 Engineering Drive 1
(Off Kent Ridge Crescent, NUS)
Singapore 117608
DID: 65-6874 8410
Mobile: 65-9446 8752

## INTRODUCTION

Spintronics[1,2] is an interesting area of research that straddles the border between fundamental physics and technology, offering an almost unique opportunity to translate physics into real applications. One example is the giant magnetoresistance[3,4] (GMR) effects which have found applications in the form of a multilayer spin valve[5,6] recording heads for reading magnetic data stored on the magnetic media. Another phenomenon, the spin transfer torque[7-9], is currently under intense investigation for current-induced magnetization switching, noise control in spin valve, and sustained spin torque oscillations[10]. Micromagnetic studies of magnetization configurations have improved the design of magnetic media and read-heads for recording purposes. Modern interest in micromagnetics consider the dynamics of both itinerant and local electron spin, providing new insights into anomalous Hall, as well as spin transfer with respect to the itinerant, local spin dynamics, respectively. Of interest recently is the spin orbital (SO) effect[11-14], particularly the Rashba and the Dresselhaus in semiconductor materials. The SO effect has direct implication to the spin and the momentum dynamics of electrons, leading to recent interest that spans fundamental, device and engineering physics, and subjects that range from spin Hall[15-17] to spin current[18] transistor. In fact, SO effect is highly relevant to spintronics, ranging from the well-known anisotropic magnetoresistance, the anisotropy energy of local moment density, the keenly studied spin Hall and spin current in semiconductor spintronics, to more subtle implications like spin torque, spin dynamics, spin oscillations, and Zitterbewegung.

A SO system can be viewed as one which provides an effective Zeeman magnetic (b) field which varies in the momentum (K) space. As such, one can draw an analogy between a SO system with a locally varying b field system, in which a conduction electron experiences the varying b field in real space (R). In the event that the electron spin evolves and aligns adiabatically to the Zeeman b field in their respective K or R spaces, under the theoretical framework of gauge and symmetry[19,20], the two systems will be analogous in that electron spin evolving adiabatically in both systems "see" a Dirac monopole in the Zeeman field space (B). Since monopole in B space has no direct bearing on the spin or orbital dynamics of electrons, it would be essential to transform the B space monopole to some topological magnetic fields (curvature) in the more useful space of K or R under which the equation of motion[15,21] can be constructed to describe the electron's orbital dynamics. A similar SU (2) system which resembles the SO is the special carbon system of monolayer and bilayer graphene. But the spinor of these systems does not represent the spin state of conducting particle in the carbon system. Instead, the spinor describes the pseudo-spin which consists of a linear combination of waves due to different sub-lattice sites. This article would be devoted to discussing the physics of monopole fields[22] originating from spinor dynamics in spintronics and graphene. The monopole field in these so-called SU(2) systems can be viewed as a mathematical object which can lead to instructive description of the electron's orbital dynamics[23,24] or motion. Here we will present a thorough description of the Dirac gauge potential arising from spinor dynamics (fast alignment with b fields) in the strong Zeeman field (adiabatic) limit. The strong Zeeman effect has direct relevance to both the SO or graphene systems and the local magnetic system; the former would relate to the transformed topological magnetic fields in K space, the latter to R space. In local micromagnetic systems which have been studied intensively in the magnetic media for hard disk drives, or domain wall spintronics, one needs to investigate the topological magnetic field in real spaces which will not be discussed in this article.

**THEORY**

When an electron propagates in the SU(2) system, its spin precesses about the effective b field. This mechanism has been studied in great details in spintronics where precession of spin due to the Rashba or the Dresselhaus effects leads to spin current[25,26] when a finite dimension (boundary condition) is imposed on the system. But here we will consider a system where the effective Zeeman b field is infinitely strong, such that in this limit electron spin relaxes to the field. The alignment of the electron spin to the local field means that the electron assumes the low-energy spin eigenstate of the system, with no admixture from the other spin eigenstate. Such system is known as adiabatic where spin is constantly aligned to the local field and there is no probability of the spin assuming its other eigenstate. One can now apply a continuous unitary transformation to the Hamiltonian such that the spin reference axis (z) in the rotated frame coincides with the local b field direction. The Hamiltonian of such system, when read in this rotated frame reveals a momentum term that appears modified. In the language of symmetry, the Hamiltonian has been transformed and corresponding transformation of the wave-function would be required to ensure the invariance of the Lagrangian.

I. *Adiabatic Gauge and Path Integral (Monopole in B space)*

As is well known, the adiabatic change of an eigenstate (eg. spin state) in parameter space (eg. b field) gives rise to a geometric phase known as the Berry's phase[27,28], which by requirement of symmetry transforms the Hamiltonian by simply modifying the momenta with a set of gauge potentials which are Abelian by nature; the general expression for the Dirac gauge potential is:

$$\tilde{A}^N = \frac{\hbar}{2e}(1-\cos\theta^N)\frac{\partial\emptyset^N}{\partial\tilde{n}}$$

(1)

where N indicates the type of space, while n is the parameter in N space. In the above, the eigenstate at any point in time has the same angular orientation $(\theta,\emptyset)$ as those of the parameters. In the event where the parameter is oriented differently than the evolving state, e.g. spin is perpendicular to the momentum in the SO system at any one time, transformation would be required to find the gauge field in the momentum space. Explicitly, Eq. (1) is $\tilde{A}^N = \frac{\hbar}{2e}\left(1-\frac{n_z}{n}\right)\left(\frac{\partial\emptyset}{\partial n_x},\frac{\partial\emptyset}{\partial n_y},0\right)$. To understand the acquisition of geometric phase in an adiabatic spinor system, we will first examine the evolution of an eigenstate from initial to final state as described by the path integral of spatial propagators as below

$$\psi(x_{n+1},t_{n+1}) = \int G(x_{n+1}t_{n+1},x_0 t_0)\,\psi(x_0,t_0)dx_0$$

(2)

where $G(x_{n+1}t_{n+1},x_0 t_0)$ is the propagator between times $t_0$ and $t$. The propagator in explicit spatial terms is

$$G(x_{n+1}t_{n+1},x_0 t_0) = \int \langle x_{n+1}|U(t_{n+1}t_n)|x_n\rangle \langle x_n|U(t_n t_{n-1})|x_{n-1}\rangle \ldots \langle x_1|U(tt_0)|x_0\rangle\, dx_1 dx_2 \ldots dx_n$$

(3)

where $\langle x_n|U(t_n t_{n-1})|x_{n-1}\rangle = \sqrt{\frac{m}{2\pi i\hbar\Delta t}} e^{\frac{im}{2\hbar}\left(\frac{x_n-x_{n-1}}{\Delta t}\right)^2\Delta t}\cdot e^{-\frac{iV}{\hbar}\Delta t} \equiv \sqrt{\frac{m}{2\pi i\hbar\Delta t}}\gamma_{n-1}$ is the propagator between two spatial points. Substituting this into Eq. (3) yields $\psi(x_{n+1},t_{n+1}) = \int \left(\frac{m}{2\pi i\hbar\Delta t}\right)^{\frac{n+1}{2}} [\gamma_n \ldots \gamma_0\, dx_n \ldots dx_1]\,\psi(x_0 t_0)\,dx_0$, which can be expressed in the simple form of:

$$\psi(x_{n+1},t_{n+1}) = \int\left[\left(\frac{m}{2\pi i\hbar\Delta t}\right)^{\frac{n+1}{2}} e^{\frac{iS(t)}{\hbar}}\, dx_n \ldots dx_1 dx_0\right]\psi(x_0 t_0)$$

(4)

where $\gamma_n \ldots \gamma_0 = e^{\frac{iS(t)}{\hbar}}$ and $S(t) = \frac{i}{\hbar}\int_0^T \frac{m}{2}\left(\frac{dx}{dt}\right)^2 - V(x_n)\,dt$ would be the action of the system. Note that $\left(\frac{m}{2\pi i\hbar\Delta t}\right)^{\frac{n+1}{2}}$ has dimension of $\left(\frac{1}{l}\right)^{n+1}$ which cancels that due to the volume element $dx_n \ldots dx_1 dx_0$. In a dynamic spinor system which evolves with the changing b fields, the infinitesimal propagator corresponding to $\sqrt{\frac{m}{2\pi i\hbar \Delta t}}\gamma_n = \langle x_{n+1}|U(t_{n+1}t_n)|x_n\rangle$ is

$$C\gamma_n = \left\langle z_{n+1}\left|e^{-\frac{i}{\hbar}\delta t\,\mu\,\tilde{b}.\tilde{\sigma}}\right|z_n\right\rangle = \pm\langle z_{n+1}|z_n\rangle e^{-\frac{i}{\hbar}\delta t\,\mu\,b} = \pm e^{-\delta t\,z^+\partial_t z}e^{-\frac{i}{\hbar}\delta t\,\mu b}$$

(5)

where $n$ and $n+1$ corresponds to interval $t$ and $t+\delta t$, respectively; $\tilde{b} = b\hat{n}$ and $\hat{n}$ is the unit vector of the b field. In the above, use has been made of the approximation $\langle z_{n+1}|z_n\rangle = 1 - z_{n+1}^+(z_{n+1} - z_n) \approx 1 - z_{n+1}^+\partial_t z_n \delta t \approx e^{-i\delta t\,z_n^+\partial_t z_n}$. Propagation from $|z_n\rangle$ to $|z_{n+1}\rangle$ depends on whether $|z_n\rangle$ is parallel or anti-parallel to the b field. Letting $\hat{n}.\tilde{\sigma} = \sigma_n$, one could deduce that $\sigma_n|z_n\rangle = \pm|z_n\rangle$ for $|z_n\rangle$ parallel / anti-parallel to $b\,\hat{n}$, note that the direction of spin state $|z_n\rangle$ is $\langle z_n|\tilde{\sigma}|z_n\rangle = \hat{n}$. It can be deduced simply by inspection that the action of the system would be $S(t) = -\mu BT + i\hbar \int_0^T z_n^+(t)\,\partial_t z_n(t)\,dt$. Neglecting the dynamic phase of $-\mu BT$ and expanding the action leads to:

$$S = \pm\frac{\hbar}{2}\int_0^t (1-\cos\theta)\frac{\partial\emptyset}{\partial t}\,dt = \pm\frac{\hbar}{2}\int_0^n (1-\cos\theta)\frac{\partial\emptyset}{\partial\tilde{b}}.d\tilde{b}$$

(6)

where $\pm$ corresponds to the parallel / anti-parallel case, respectively. Here, $(\theta,\emptyset)$ is understood to be $(\theta^B,\emptyset^B)$. The above is the Berry's phase which can be associated with a gauge field in B space for the system under consideration. The term $\frac{\hbar}{2}(1-\cos\theta)\frac{\partial\emptyset}{\partial\tilde{b}}$ is a gauge potential defined on the $S_2$ manifold as a regular expression except at $\theta = \pi$. Thus the curvature of this term does not represent a regular quantity defined everywhere on a $S_2$ manifold which is parametrized by $(r,\theta,\emptyset)$; in fact it has a singularity on the $-z$ axis (see Appendix). This problem was resolved by Wu and Yang[29,30] by conceiving that the gauge over the $S_2$ should be represented by at least two different expressions, each expression covers any part of the manifold but avoids its own singularity. The overlap between the two expressions is the area where transition from one expression to the other must be carried out. In this manner, a $\oint (\tilde{\nabla}_B \times \tilde{A}^B).d\tilde{S}$ which relies on two cross sections of the gauge potential functions will be non-vanishing, indeed it can be proven to be a monopole field. In summary, it has been shown that the path integral of the spinor dynamic in B space under adiabatic approximation has thus generated a Dirac potential and consequently a Dirac monopole field which can recently be treated as the more generalized Wu-Yang monopole. But the monopole field obtained in the B space is not instructive with respect to interpreting its physical effect on the dynamics of spin or charged particle. Transformation of the gauge potential in B space to the more relevant K or R space will thus be crucial for providing a more insightful understanding. We will now focus on $+\frac{\hbar}{2}(1-\cos\theta)\frac{\partial\emptyset}{\partial \tilde{b}}$ which arises due to the spin assuming only the lower energy eigenstate of the magnetic fields, and discuss the transformation of this gauge potential from one space to another. The gauge field in N space can be converted to an arbitrary L space by

$$A_\mu^L = \frac{\hbar}{2e}(1-\cos\theta)\frac{\partial\emptyset}{\partial\tilde{n}}\cdot\frac{\partial\tilde{n}}{\partial l_\mu} \equiv \tilde{A}^N\cdot\frac{\partial\tilde{n}}{\partial l_\mu}$$

(8)

We have shown that Eq. (8) is a result of adiabatic evolution of a spin up particle tracking the b field. Alternatively, one could use the rotation matrix to perform a local transformation of the Hamiltonian, resulting in $1/2m\,(p_\mu - i\hbar U\partial_\mu U^+)^2$. It is then worth noting that $\tilde{A}^N\cdot\frac{\partial\tilde{n}}{\partial l_\mu}$ constitutes the top left diagonal term of $U\partial_\mu U^+$ where U is the unitary matrix which rotates the laboratory axis to the b field where spin is aligned to. The above descriptions can also be summarized by $\tilde{A}^N\cdot\frac{\partial\tilde{n}}{\partial l_\mu} = diag[U\partial_\mu U^+] = A_{mon}^v \partial_\mu n^v$, where it is also common to write $A_{mon}^v = A(\theta)\,\partial\emptyset/\partial n^v$, which is none other than the monopole field in N space. In fact, it is legitimate to ask if the Dirac gauge potential above has any physical implication to a particle which experiences its presence because such Dirac potential is not well-defined ($\theta = \pi$) everywhere. Its surface integral is vanishing as a consequence. This suggests that the Dirac potential derived through the path integral approach is mathematically inadequate to represent the monopole field. To make a case for the existence of the monopole field, the Dirac string $\delta(x)\delta(y)\theta(-z)$ must be avoided. It becomes clear that negotiating the Dirac string and covering the manifold completely is required to obtain a curvature field which resembles a physical magnetic field. In applied physics, one can thus view the Wu Yang's treatment of the monopole field as having reasonably affirmed that the monopole field here can be regarded as a physical magnetic field which influences the dynamics of charged particle in the same way that physical magnetic field does. One can by now reason that adiabatic spinor evolution gives rise to a monopole field, which might have significant physical implications to particles which experience its presence. Since we know that the Dirac string can be negotiated, we will not show the Dirac string in the derivation of the monopole field (Dirac string will be covered in the Appendix). The curvature of the gauge potential in B space, which according to the electrodynamics tensor is

$$\Omega_\kappa^B = \frac{\partial}{\partial b_\mu} A_\nu^B \varepsilon_{\mu\nu\kappa}$$

(9)

where $\mu, \nu, \kappa$ are the 3 spatial dimensions which sum over double index by Einstein's convention. Noting the explicit form of $\tilde{A}^B$, and setting $\hbar/2e$ to 1, one obtains

$$\Omega_\kappa^B = \left(\left(1 - \frac{b_z}{b}\right)\frac{\partial^2 \emptyset}{\partial b_\mu \partial b_\nu} - \left(\frac{\partial}{\partial b_\mu}\frac{b_z}{b}\right)\frac{\partial \emptyset}{\partial b_\nu}\right)\varepsilon_{\mu\nu\kappa}$$

(10)

For clarity, the z component of the curvature can be deduced directly from Eq. (9) to be

$$\Omega_z^B = \left(1 - \frac{b_z}{b}\right)\left[\frac{\partial}{\partial b_x}, \frac{\partial}{\partial b_y}\right]\emptyset - \left(\frac{\partial}{\partial b_x}\frac{b_z}{b}\right)\frac{\partial \emptyset}{\partial b_y} + \left(\frac{\partial}{\partial b_y}\frac{b_z}{b}\right)\frac{\partial \emptyset}{\partial b_x}$$

(11)

The monopole field has been expressed explicitly as a function of the Zeeman b field. In SU(2) system, it is normally straightforward to deduce an effective Zeeman b field from the Hamiltonian. The Zeeman b field would be related to the energy which forms the generator of time translation for the spinor part of the wavefunction. To derive the curvature in B space explicitly, we note that $\cos\theta = b_z/b$, $\tan\phi = b_y/b_x$; restoring $\hbar/2e$, the curvature reveals their monopole signatures:

$$\Omega_x^B = \frac{\hbar}{2e}\frac{b_x}{b^3}, \qquad \Omega_y^B = \frac{\hbar}{2e}\frac{b_y}{b^3}, \qquad \Omega_z^B = \frac{\hbar}{2e}\frac{b_z}{b^3}$$

(12)

Note that in the above expressions, the Dirac string has been deliberately ignored; more discussion of the Dirac string will be given in the Appendix. One could easily check that had the above derivation been carried out in real space, i.e. replacing b with r, expressions of Eq. (12) would have the dimension of the magnetic fields. However, in application this is only possible for the specific case of spin aligning with r, i.e. spin has the same orientation as the r coordinates with respect to some spherical center. One way of achieving this is by means of technology, create magnetic fields or local magnetic moments with orientations in B space that overlaps the r coordinates in R space with respect to spherical centers in the resepective spaces. We will not discuss this in details here, the focus of this paper will be the spin orbital or the graphene system where topological magnetic field (curvature) could exist naturally in the K space.

## II. Topological Magnetic Fields in K (reciprocal) space in various SU(2) systems

In the above, we have shown the path integral derivation of the gauge potential and its monopole in the B space, as a result of adiabatic spin alignment with the b fields. But the monopole curvature in b field is not useful for elucidating even heuristically the orbital dynamics of a particle which "sees" these fields. In nature, there exist many systems that provide the mimic of Zeeman b fields, and these fields normally depend on the momentum (k). It is of interest indeed to find the topological magnetic field in the K space which would be more instructive for deriving the equation of motion[15,21], hence for the elucidation of orbital dynamics of particles which "see" these fields. The SO systems, which have long been known to exist in atomic physics (hyperfine interaction), semiconductors, metals, are the most conspicuous systems which provide such k-dependent Zeeman fields. SO effect manifests in the bandstructure, lifting degeneracy of valence electrons at momentum other than zero. Other b(k) systems include the carbon based systems eg. monolayer and bilayer graphene, and superconducting systems. The formal description of SO as well as graphene pseudo spin effect could be derived using the Dirac's formalism which suggests that these effects mimic the vacuum SO effect in the non-relativistic limit[28,29]; Dirac matrix equations are shown below

$$\begin{pmatrix} \left(\frac{\mathcal{E}}{c} - e\frac{\phi}{c}\right) - mc & -\tilde{\sigma}.\tilde{p} \\ \tilde{\sigma}.\tilde{p} & -\left(\frac{\mathcal{E}}{c} - e\frac{\phi}{c}\right) - mc \end{pmatrix} \cdot \begin{pmatrix} \psi \\ \chi \end{pmatrix} = 0 \tag{13}$$

$$\begin{pmatrix} (\mathcal{E} - e\phi)^2 - c^2 p^2 - m^2 c^4 & -ie\hbar c\,\tilde{\sigma}.\tilde{E} \\ -ie\hbar c\,\tilde{\sigma}.\tilde{E} & (\mathcal{E} - e\phi)^2 - c^2 p^2 - m^2 c^4 \end{pmatrix} \cdot \begin{pmatrix} \psi \\ \chi \end{pmatrix} = 0 \tag{14}$$

Using the relation $(\tilde{\sigma}.\tilde{E})(\tilde{\sigma}.\tilde{p}) = i\sigma.(\tilde{E} \times \tilde{p}) + \tilde{E}.\tilde{p}$, and solving for $\psi$, one obtains

$$\left((\mathcal{E} - e\phi)^2 - c^2 p^2 - m^2 c^4 - \frac{ie\hbar c^2 (\tilde{E}.\tilde{p})}{\mathcal{E} - e\phi + mc^2} + e\hbar c^2 \frac{\sigma.(\tilde{E} \times \tilde{p})}{\mathcal{E} - e\phi + mc^2}\right)\psi = 0 \tag{15}$$

In the non-relativistic limit, i.e. assuming a small $\epsilon = \mathcal{E} - mc^2$ and $e\phi$, the above reduces to the following standard Hamiltonian for SO coupling

$$\left(\frac{p^2}{2m} + e\phi + \frac{\hbar e}{4m^2 c^2}\sigma.(p \times E) + \frac{i\hbar e}{4m^2 c^2}E.p\right)\psi = \epsilon \psi \tag{16}$$

In application to particles in graphene like systems which mimic Dirac fermions due to material bandstructure, it would be instructive to replace the coupling mass term of $mc^2$ for particles in vacuum with a coupling term $\Delta$ which arises due to material bandstructure but plays the same role as the mass term, as far as the Dirac matrix is concerned. The coupling term $\Delta$ gives rise to the energy dispersion where the effective mass of particles in the materials can be derived; in other words, particle effective mass is a function of $\Delta$ but not vice versa. For monolayer graphene, $\Delta$ vanishes and it can be derived from the energy dispersion relation that particles behave like massless Dirac fermions. In semiconductor physics, SO effects have been known to exist both in the conduction and the valence band; its effect on the bandstructure has been well studied. In the technology relevant spintronics, the more specialized type of Rashba and Dresselhaus SO coupling have been considered for generating spin current in semiconductor spintronics. SO effects which manifestly appear in the energy dispersion relation often results in a coupled form of spin-dependent transmission of electrons across potential barriers. Equation (16) shows that the SO system provides a b(k) relation via $\sigma.(p \times E)$ which is crucial for the derivation of the curvature in K space. The other systems which give the required b(k) relation is the graphene system. The b(k) in these systems originate from the bandstructure which exhibits a linear energy dispersion around the two inequivalent, hexagonal corners (valley) of the first Brilliouin zone, also known as the Fermi points. The prevalence of b(k) systems in nature motivate us to present a systematic approach to study the gauge potential and its topological curvature in the K space. In SO, graphene or superconducting systems, the parameter of interest lie in the K space. One recalls that in the system where the spinor evolves and aligns to the b fields, the gauge fields in B and K spaces, are respectively, $\tilde{A}^B = \frac{\hbar}{2e}(1 - \cos\theta^B)\frac{\partial \phi^B}{\partial \tilde{b}}$ and $\tilde{A}^K = \frac{\hbar}{2e}(1 - \cos\theta^B)\frac{\partial \phi^B}{\partial \tilde{k}}$; the two expressions are related by $A_\mu^K \equiv \tilde{A}^B . \frac{\partial \tilde{b}}{\partial k_\mu}$. To find the Dirac gauge potential in K space, one merely needs b as a function of k. Space conversion for the gauge potential is straightforward, but that for the monopole field will be slightly more complex. But we will show below that in fact to transform the monopole field in B space to a topological curvature in K space, it also suffices to have the Zeeman b as a function of k. Looking at the z component only,

$$\Omega_z^K = \frac{\partial A_y^K}{\partial k_x} - \frac{\partial A_x^K}{\partial k_y}$$

(17)

Using $A_\mu^K \equiv \tilde{A}^B . \frac{\partial \tilde{b}}{\partial k_\mu}$, it is straightforward to obtain

$$\Omega_z^K = \frac{\partial}{\partial k_x}\left(\tilde{A}^B \cdot \frac{\partial \tilde{b}}{\partial k_y}\right) - \frac{\partial}{\partial k_y}\left(\tilde{A}^B \cdot \frac{\partial \tilde{b}}{\partial k_x}\right) = \left(\frac{\partial \tilde{b}}{\partial k_x} \cdot \partial_{\tilde{b}}\right)\left(\tilde{A}^B \cdot \frac{\partial \tilde{b}}{\partial k_y}\right) - \left(\frac{\partial \tilde{b}}{\partial k_y} \cdot \partial_{\tilde{b}}\right)\left(\tilde{A}^B \cdot \frac{\partial \tilde{b}}{\partial k_x}\right)$$

(18)

which in tensor form is

$$\Omega_z^K = \left(\frac{\partial b_\nu}{\partial k_x} \frac{\partial A_\mu^B}{\partial b_\nu} \frac{\partial b_\mu}{\partial k_y}\right) - \left(\frac{\partial b_\nu}{\partial k_y} \frac{\partial A_\mu^B}{\partial b_\nu} \frac{\partial b_\mu}{\partial k_x}\right) + A_\mu^B\left(\frac{\partial b_\nu}{\partial k_x} \frac{\partial^2 b_\mu}{\partial b_\nu \partial k_y} - \frac{\partial b_\nu}{\partial k_y} \frac{\partial^2 b_\mu}{\partial b_\nu \partial k_x}\right)$$

(19)

The last term written in $A_\mu^B\left(\frac{\partial^2 b_\mu}{\partial k_x \partial k_y} - \frac{\partial^2 b_\mu}{\partial k_y \partial k_x}\right)$ clearly vanishes as $[\partial_x, \partial_y] = 0$ in general.

$$\Omega_z^K = \left(\frac{\partial A_\mu^B}{\partial b_\nu} - \frac{\partial A_\nu^B}{\partial b_\mu}\right) \frac{\partial b_\nu}{\partial k_x} \frac{\partial b_\mu}{\partial k_y} = \frac{\partial A_i^B}{\partial b_j}(\delta_{i\mu}\delta_{j\nu} - \delta_{i\nu}\delta_{j\mu})\left(\frac{\partial b_\nu}{\partial k_x} \frac{\partial b_\mu}{\partial k_y}\right)$$

(20)

Using the identity $(\delta_{i\mu}\delta_{j\nu} - \delta_{i\nu}\delta_{j\mu}) = \varepsilon_{ijk}\varepsilon_{\mu\nu k}$ where $\varepsilon_{ijk}$ is the fully anti-symmetric tensor, $\delta_{i\mu}$ is the kronecker delta, one gets the the z-component of the curvature in K space

$$\Omega_z^K = \frac{\partial A_i^B}{\partial b_j}\varepsilon_{ijk}\left(\frac{\partial b_\nu}{\partial k_x} \frac{\partial b_\mu}{\partial k_y}\right)\varepsilon_{\mu\nu k}$$

(21)

Likewise, the other components of this curvature field can be derived, and expressed in the more familiar vector calculus form:

$$\Omega_x^K = \frac{\tilde{b}}{b^3} \cdot \left(\frac{\partial \tilde{b}}{\partial k_z} \times \frac{\partial \tilde{b}}{\partial k_y}\right)$$

$$\Omega_y^K = \frac{\tilde{b}}{b^3} \cdot \left(\frac{\partial \tilde{b}}{\partial k_x} \times \frac{\partial \tilde{b}}{\partial k_z}\right)$$

$$\Omega_z^K = \frac{\tilde{b}}{b^3} \cdot \left(\frac{\partial \tilde{b}}{\partial k_y} \times \frac{\partial \tilde{b}}{\partial k_x}\right)$$

(22)

where $\frac{\partial A_i^B}{\partial b_j}\varepsilon_{ijk} = \frac{\tilde{b}}{b^3}$. It has thus been shown that the topological curvature in K space can be developed by substituting the Zeeman b field into Eq. (22). It is of interest to remind readers that had $\tilde{b} = c\tilde{k}$ or $\tilde{b} = c\tilde{k}$, no conversion is required, and the monopoles will, respectively be $\Omega_\mu^K = \frac{\hbar}{c2e}\frac{\tilde{k}}{k^3}$ or $\Omega_\mu^R = \frac{\hbar}{c2e}\frac{\tilde{r}}{r^3}$ when one follows the Eqs.(9)-(12) in their respective spaces. When the spin of a particle inscribes a path on $S_B^2$, the gauge potential it "sees" at every point on the path is $A^B = \frac{\hbar}{2e}(1 - \cos\theta^B)d\emptyset^B$. However, when a particle inscribes a path on $S_K^2$, the gauge it "sees" is $A^K \neq \frac{\hbar}{2e}(1 - \cos\theta^K)d\emptyset^K$ unless the relation $\tilde{b} = c\tilde{k}$ is obeyed. This is because $A = \frac{\hbar}{2e}(1 - \cos\theta^B)d\emptyset^B$ is obtained as a result of spin assuming the eigenstate of $\tilde{b}(t)$ in an adiabatic manner, the spin thus shares the same polar coordinates with the b field. To obtain the curvature in K space, one needs to convert $A^B$ to and then perform $\nabla_K \times \tilde{A}^K$. Since $A^K \neq \frac{\hbar}{2e}(1 - \cos\theta^K)d\emptyset^K$, it is worth cautioning that while $\nabla_B \times \tilde{A}^B$ is a monopole field, $\nabla_K \times \tilde{A}^K$ might not be one. It is important to note that curvature in K space cannot be obtained by substituting the effective b fields into Eqs. (12); doing this is merely re-expressing $\nabla_B \times \tilde{A}^B$ in B space as a function of k., but not performing the function of Eqs. (22). Equations (22) are most useful in SO or graphene systems, which comprise k-dependent Zeeman b field in the Lorentz frame of a charge particle, where k is the electron wavevector.

Below, we provide the topological curvature in K space for the cubic Dresselhaus, the linear Dresselhaus, the Perel-modified Dresselhaus, the linear Rashba and Dresselhaus in quantum well systems, as well as the monolayer Weyl system of massless Dirac fermions in graphene. We present below the various SO effects and their corresponding effective b fields. Effective Zeeman b fields can be derived by considering the first of $H = \mu_B b = \frac{eg\hbar}{4m}b$ where $\mu_B$ is the Bohr magneton that has the SI unit of Joule/Tesla; note that $\mu_B$ is also equivalent to $IA$ where $I$ is a circulating current and $A$ is the area enclosed by the circulating current,

Table I. Reference of various spin orbital systems, their Hamiltonian, relativistic effective fields, and corresponding K-space monopole curvature.

| Spin Orbit Coupling Types (Material System) | Hamiltonian and Effective b fields | $\tilde{\Omega}$ is the curvature field in K-space; |
|---|---|---|
| Semiconductor systems with linear SO coupling (GaAs, GaSb, InAs, InSb, GaInAs, GaInSb, ) ||| 
| Linear Dresselhaus | $H = \eta_D(\sigma_x k_x - \sigma_y k_y)$  $\tilde{b} = \frac{\eta_D}{\mu_B}\begin{pmatrix} k_x \\ -k_y \\ 0 \end{pmatrix}$ | $\tilde{\Omega} = \pm\frac{\hbar\pi}{e}\begin{pmatrix} 0 \\ 0 \\ \delta^2(k) \end{pmatrix}$ |
| Linear Rashba | $H = \eta_R(\sigma_x k_y - \sigma_y k_x)$ $$\tilde{b} = \frac{\eta_D}{\mu_B}\begin{pmatrix} k_y \\ -k_x \\ 0 \end{pmatrix}$$ | $$\tilde{\Omega} = \pm \frac{\hbar\pi}{e}\begin{pmatrix} 0 \\ 0 \\ \delta^2(k) \end{pmatrix}$$ |
| Graphene systems | | |
| Monolayer graphene (Massless Weyl) | $H = A(\sigma_x k_x + \sigma_y k_y)$ $$\tilde{b} = \frac{A}{\mu_B}\begin{pmatrix} k_x \\ k_y \\ 0 \end{pmatrix}$$ | $$\tilde{\Omega} = \pm \frac{\hbar\pi}{e}\begin{pmatrix} 0 \\ 0 \\ \delta^2(k) \end{pmatrix}$$ |
| Bilayer graphene | $$H = \frac{\hbar^2}{2m}\left(\sigma_x \frac{k_+^2 + k_-^2}{2} - \sigma_y \frac{k_-^2 - k_+^2}{2i}\right)$$ $$\tilde{b} = \frac{1}{2\mu_B m}\begin{pmatrix} k_x^2 - k_y^2 \\ 2k_x k_y \\ 0 \end{pmatrix}$$ | $$\tilde{\Omega} = \pm \frac{\hbar 2\pi}{e}\begin{pmatrix} 0 \\ 0 \\ \delta^2(k) \end{pmatrix}$$ |
| Semiconductor systems with cubic SO coupling | | |
| Dresselhaus Cubic (Bulk III-V) | $H = \eta_{DC}\left(\sigma_x k_x [k_y^2 - k_z^2]\right) + c.p.$ $$\tilde{b} = \frac{\eta_{DC}}{\mu_B}\begin{pmatrix} k_x[k_y^2 - k_z^2] \\ k_y[k_z^2 - k_x^2] \\ k_z[k_x^2 - k_y^2] \end{pmatrix}$$ | $$\tilde{\Omega} = \frac{\pm\hbar}{2eb^3}\left((k_x^2 - k_y^2) + cp\right)\begin{pmatrix} k_x \\ k_y \\ k_z \end{pmatrix}$$ |
| Dresselhaus (high kinetic, Perel) | $H = \eta_{Dk}(-\sigma_x k_x k_z^2 + \sigma_y k_y k_z^2 + \sigma_z k_z[k_x^2 - k_y^2])$ $$\tilde{b} = \frac{\eta_{Dk}}{\mu_B}\begin{pmatrix} -k_x k_z^2 \\ k_y k_z^2 \\ k_z[k_x^2 - k_y^2] \end{pmatrix}$$ | $$\tilde{\Omega} = \frac{\pm\hbar}{2eb^3}(k_y^2 - k_x^2)\begin{pmatrix} k_z^4 k_x \\ k_z^4 k_y \\ k_z^5 \end{pmatrix}$$ |

Heuristically, the topological magnetic field ($\Omega$) can be regarded as the source providing Lorentz forces that affect electron motion. The fact that the force experienced by the spin up /down particle is opposite in directions arising due to $\pm(\nabla \times \tilde{A})$ provides the physical picture of spin transverse separation. However, more rigorous quantification of such separation, or the Hall conductivity would have to come from the Berry's phase and the Chern-Simon term which can be interpreted as a summation of the Lorentz forces over all electron momenta. The Berry's phase in K space which has important relevance to electron transport can be derived from $\emptyset = \frac{e}{\hbar}\int \tilde{A}.d\tilde{k} = \frac{e}{\hbar}\int (\nabla \times \tilde{A}).d^2\tilde{k}$, where it would be worth noting that $(\nabla \times \tilde{A})$ can be viewed as the topological magnetic field that the electron "feels". It would not be the focus of this paper to conduct extended analysis of the Berry's phase of all systems. But in view of modern trend of quantifying system conductivity using the Berry's phase, we merely provide its derivations for common systems only, i.e. linear Rashba, linear Dresselhaus, monolayer and bilayer graphene. To derive the Dirac potential for the above systems in the intended space of K, one needs to recall that $A_\mu^k = \frac{\hbar}{2e}(1-\cos\theta)\frac{\partial \emptyset}{\partial k_\mu}$ but noting that $\emptyset$ and $\theta$ are angles defined by the B field. To derive $A_\mu^k$ for the systems above, one would thus need the effective magnetic fields of each system. Keeping in mind that in the B space, $\cos\theta = b_z/b$, $\tan\emptyset = b_y/b_x$, one derives $A_\mu^k$. To obtain the Berry's phase, let us examine the gauge of $\tilde{A} = \pm\frac{\hbar}{2ek^2}\begin{pmatrix}pk_y \\ qk_x \\ 0\end{pmatrix}$ which is the general expression for the above-mentioned systems at $\theta = \pi/2$, i.e. in the plane of the R as well as K space two-dimensional (2D) system. The Berry's phase can be derived with $\oint_C \tilde{A}\left(\theta = \frac{\pi}{2}\right).d\tilde{k}$ where C represents a closed path in the plane surrounding the monopole at $(k_x, k_y) = 0$. Alternatively, one could perform by Stoke's theorem $\oint_{SN}[\nabla \times \tilde{A}(\theta,\emptyset)].d^2\tilde{k}$ where SN represents the entire curved surface of the North hemisphere of the $S^2$ manifold. We can equivalently evaluate $\oint_{SF}\left[\nabla \times \tilde{A}\left(\theta = \frac{\pi}{2}\right)\right].d^2\tilde{k}$ on SF which represents the flat circular surface bound by C. The results are summarized in Table II for different systems corresponding to different values of $p, q$.

Table II. Reference of various spin orbital systems, their gauge, and corresponding Berry's phase.

| Spin Orbit Coupling Types (Material System) | Gauge | $\emptyset$ is the Berry's phase $\emptyset = \frac{e}{\hbar}\int \tilde{A}.d\tilde{k} = \oint_{SF}\left[\nabla \times \tilde{A}\left(\theta = \frac{\pi}{2}\right)\right].d^2\tilde{k}$ |
|---|---|---|

| Linear Dresselhaus | $\tilde{A} = \pm \dfrac{\hbar}{2ek^2}\begin{pmatrix} k_y \\ -k_x \\ 0 \end{pmatrix}$ | $n\pi$ |
|---|---|---|
| Linear Rashba | $\tilde{A} = \pm \dfrac{\hbar}{2ek^2}\begin{pmatrix} -k_y \\ k_x \\ 0 \end{pmatrix}$ | $n\pi$ |
| Massless Weyl (monolayer graphene) | $\tilde{A} = \pm \dfrac{\hbar}{2ek^2}\begin{pmatrix} -k_y \\ k_x \\ 0 \end{pmatrix}$ | $n\pi$ |
| Massive Dirac (bilayer graphene) | $\tilde{A} = \pm \dfrac{\hbar}{2ek^2}\begin{pmatrix} -2k_y \\ 2k_x \\ 0 \end{pmatrix}$ | $2n\pi$ |

We have introduced a general approach which allows one to derive the monopole field of a specific system in the B space using the Dirac gauge potential which is defined in the same B space. We then focused on the various spintronic and graphene systems in which such monopole fields are relevant. In the SO systems relevant to semiconductors as well as graphene systems where the Zeeman b fields are k-dependent, we showed how the topological curvature (magnetic field) in the more useful K space can be derived. The physical significance of the curvature fields depends on the space in which the curvature is taken. The usefulness of the approach in this paper is that it provides a unified underlying picture for the curvature fields any arbitrary spaces (e.g. in K and R spaces) under a common origin, i.e. the Dirac gauge potential and its monopole field in B space. One merely requires an effective Zeeman b field of the form $\sigma.B$ in the Hamiltonian of a specific system for a quick derivation of the topological curvature in any space outside of the B space.

The surface integral of the curvature yields a non-vanishing quantized value, which is invariant under deformation of the surface of integration; it is hence a topological object. In the context of Dirac monopole, this is associated with the quantization of the electric charge. In the SO or graphene systems, this quantity is associated with quantized magnetic flux or the Berry's phase. It therefore becomes clear that the existence of K-space topological curvature (magnetic field) can be related heuristically to particle trajectory, and in the case of SO system, spin-dependent separation of charges would be resulted from the spin-dependent curvature fields in K space. Our summary in Table I provides a unifying picture for the physics of particle motion in various systems eg. spin orbital, local magnetic, graphene and superconducting systems under the theory of symmetry, gauge and monopoles. The approach and results of our work allow experimentalists to seek out material system which exhibit Zeeman-like terms for potential spin Hall, quantum spin Hall, SO-induced spin torque, spin oscillation, and other measurements. In modern treatment, Berry's phase has been associated with quantized conductivity of mesoscopic systems. Table II thus provides a unifying picture of the various types of quantum Hall effects that could possibly exist in different condensed matter systems.

**APPENDIX (Topological Properties of the Monopole Fields)**

To determine the vector potential for a regular monopole field is not a trivial task. In fact a single vector potential function which is regular everywhere on the $S_2$ manifold probably does not exist. We would like to remind readers that a $(\widetilde{\nabla}_B \times \tilde{A}^B)$ based on one gauge expression (either North or South pole) will necessarily yield a vanishing surface integral over the entire $S^2$ manifold, due to the Dirac string. Whereas in modern understanding where $(\widetilde{\nabla}_B \times \tilde{A}^B)$ is based on at least two gauge expressions, the curvature is regular everywhere even though the individual gauge expression is not, and the surface integral of such regular curvature will be a non-vanishing quantity as the string can be avoided. Here we will show using the South Pole gauge that the Dirac magnetic field has a string. The vector form of the Dirac potential is:

$$\tilde{A}^S = \frac{\hbar(\tilde{b} \times \tilde{k})}{2eb\left(b - \tilde{b}.\tilde{k}\right)} = \frac{\hbar \; b_y i - b_x j}{2eb \; (b - b_z)}$$

(23)

At first sight, the curvature of the above potential may appear to be $(\widetilde{\nabla}_B \times \tilde{A}^S) = \frac{\hbar \tilde{b}}{2eb^3}$. However, the expression is incomplete because it does not fully capture the fact. The Dirac potential is not a regular function; it is singular along $\theta = 0$ but regular along $\theta = \pi$. A proper approach is to first regularize the Dirac potential by inserting a $\epsilon^2$ to R so that $\theta = 0$ can be negotiated, and derive the regularized magnetic field. The regularized potential is

$$\tilde{A}^S = \frac{\hbar(\tilde{b} \times \tilde{k})}{2e \; R \; \left(R - \tilde{b}.\tilde{k}\right)}$$

(24)

where $R^2 = b_x^2 + b_y^2 + b_z^2 + \epsilon^2$. The regularized magnetic field can be shown to be:

$$\widetilde{\Omega}(b,\epsilon) = \frac{\hbar}{2e}\frac{\tilde{b}}{R^3} - \frac{\hbar}{2e}\left(\frac{\epsilon^2}{R^2(R-b_z)^2} + \frac{\epsilon^2}{R^3(R-b_z)}\right)\tilde{k}$$

(25)

Lifting regularization which simply means presenting a modified expression which resides in both the regular and the singular regions, one obtains:

$$\lim_{\epsilon \to 0} \widetilde{\Omega}(b,\epsilon) = \frac{\hbar}{2e}\frac{\tilde{b}}{b^3} - \frac{\hbar}{2e}4\pi\theta(b_z)\delta(b_x)\delta(b_y)$$

(26)

Equation (26) consists of two parts, the regular magnetic monopole and the singular Dirac string. The curvature of Eq. (26) vanishes as $\oint \widetilde{\Omega} \cdot d\tilde{S} = \oint \widetilde{\Omega}_r \cdot d\tilde{S} + \oint \widetilde{\Omega}_{string} \cdot d\tilde{S} = 4\pi g - 4\pi g = 0$, noting here that the monopole has been separated into the regular and the string parts. This clearly shows that the South pole Dirac potential used above cannot be a correct representation of the vector potential for a magnetic monopole. It is thus necessary to find a correct representation and this task was completed with the modern theory of fiber bundle by Wu and Yang. In this treatment, it is important to understand that there should exist at least two gauge expressions on the $S^2$ manifold in order to provide a non-vanishing $\oint (\widetilde{\nabla}_B \times \tilde{A}^B) \cdot d\tilde{S}$ over the manifold. In fact by Stokes' theorem, one sees that the surface integral of the monopole field is the line integral of the gauge field. To illustrate this more clearly, one resorts to the differential form which is related to the vector quantity as follows:

$$A = \frac{\hbar}{2e}(1-\cos\theta)\left(\frac{\partial \emptyset}{\partial \tilde{b}}\right).d\tilde{b} = \frac{\hbar}{2e}(1-\cos\theta)d\emptyset$$

(27)

$$dA = (\widetilde{\nabla} \times \tilde{A}).d\tilde{S}$$ (28)

In Eq. (27), $A$ is expressed as a 1-form. For comparison in K space, $A = \frac{\hbar}{2e}(1-\cos\theta^B)\left(\frac{\partial \emptyset^B}{\partial \tilde{b}}\right).\left(\frac{d\tilde{b}}{\partial \tilde{k}}\right).d\tilde{k}$ is consistent with $A^K \neq \frac{\hbar}{2e}(1-\cos\theta^K)d\emptyset^K$. Performing an exterior differentiation of A following standard definitions in differential geometry will lead naturally to Eq. (28) which is the dot areal product of the curvature; b is the parameter of an arbitrary space eg. the B space and dS is in that space too. According to Wu Yang, in fact Eq. (27)

is not a unique, nor is it a complete form of the gauge potential. The gauge should be expressed as one of Eq. (29) depending on the chart which covers the $S_b^2$ surface on which A is defined, b is the radius. One example set of two charts that overlap but completely cover the $S_b^2$ manifold is:

$$A^S = -\frac{\hbar}{2e}(1+\cos\theta)d\phi \qquad A^N = \frac{\hbar}{2e}(1-\cos\theta)d\phi$$

(29)

$dA^N$ is defined everywhere on $S^2$ except the –z axis, while $dA^S$ is defined everywhere except the +z axis. Equation (29) is the differential form of the Dirac gauge potential in the spherical polar coordinates. For comparison in Cartesian coordinates, the differential form of the Dirac gauge is:

$$A^S = \frac{\hbar}{2e}\left(\frac{-b_y}{b(b-b_z)}, \frac{b_x}{b(b-b_z)}, 0\right) \qquad A^N = \frac{\hbar}{2e}\left(\frac{-b_y}{b(b+b_z)}, \frac{b_x}{b(b+b_z)}, 0\right)$$

(30)

where b is the required coordinates in Cartesian form. Since $A^N$ is related to $A^S$ by $A^N = A^S + \mathrm{d}\theta$, it is obvious that $dA^S = dA^N$ which shows $dA$ is unique. In summary, $A$ is not globally defined throughout $S_b^2$, otherwise $\int dA \neq 0$; in other words, if there exists a vector $\tilde{A}$ such that its curvature has no singularity, and $\tilde{A}$ is unique, then $\int dA \neq 0$. Now with the Dirac gauge potential defined on the North and South charts, one can then perform the surface integral of these gauge potentials as $\oint \tilde{\Omega} \cdot d\tilde{S} = \oint_N (\nabla \times \tilde{A}^N) \cdot d\tilde{S} + \oint_S (\nabla \times \tilde{A}^S) \cdot d\tilde{S} = 2\pi g + 2\pi g = 4\pi g$, thus avoiding the string of each gauge potential. Although $A^N$ and $A^S$ are by no means unique on their own, they uniquely determine a 1-form $\omega$ on the bundle space P which is a $S_R^3$, where R is the radius of the 3-sphere. Here we provide the mathematical origin, the 1-form $\omega$ on $\mathbb{R}_4$ is given by:

$$\omega = i(-y_1 dx_1 + x_1 dy_1 - y_2 dx_2 + x_2 dy_2)$$

(31)

which is well-defined on the $S_R^3$. The $S_R^3$ is itself defined by $S_R^3 = \{(x_1, y_1, x_2, y_2) \in \mathbb{R}_4 : x_1^2 + y_1^2 + x_2^2 + y_2^2 = 1\}$. One can reparametrize $S_R^3$ such that $S_R^3 = \left\{(z_1, z_2) = \left(\cos\frac{\theta}{2}e^{i\varepsilon_1}, \sin\frac{\theta}{2}e^{i\varepsilon_2}\right) : 0 \leq \frac{\theta}{2} \leq \frac{\pi}{2}, \varepsilon_1, \varepsilon_2 \in \mathbb{R}_4\right\}$, while obeying $x_1^2 + y_1^2 + x_2^2 + y_2^2 = 1$. Every point of $(z_1, z_2) = \left(\cos\frac{\theta}{2}e^{i\varepsilon_1}, \sin\frac{\theta}{2}e^{i\varepsilon_2}\right) \in S_R^3$ can then be mapped via $\mathcal{P}(\theta, \varepsilon_1, \varepsilon_2) = (\sin\theta\cos(\varepsilon_1 - \varepsilon_2), \sin\theta\sin(\varepsilon_1 - \varepsilon_2), \cos\theta)$ to a point on $S_b^2$ where $\mathcal{P}: S^3 \to S^2$ which is in fact a form of Hopf mapping to project the $\omega$ on $S_R^3$ to $S_b^2$. In other words, the Hopf mapping $\eta: S_R^3 \to S_b^2$ yields a vector potential regular everywhere on $S_R^3$. It is worth noting that any point of $(z_1, z_2) \in S_R^3$ satisfies:

$$\left(|z_1|, \frac{z_2|z_1|}{z_1}\right) = \left(\cos\frac{\theta}{2}, \sin\frac{\theta}{2}e^{-i\phi}\right); \quad \left(\frac{z_1|z_2|}{z_2}, |z_2|\right) = \left(\cos\frac{\theta}{2}e^{i\phi}, \sin\frac{\theta}{2}\right),$$

(32)

The former corresponds to $(z_1, z_2) \times \frac{|z_1|}{z_1}$, the latter $(z_1, z_2) \times \frac{|z_2|}{z_2}$. The same point of $(z_1, z_2) \in S_R^3$ can be re-expressed in Eq. (32) in two different forms known as the cross section, each corresponding to one part of the $S_b^2$; $\left(\cos\frac{\theta}{2}, \sin\frac{\theta}{2}e^{-i\phi}\right)$ corresponds to $U_N = S_b^2 - (0,0,-1)$, $\left(\cos\frac{\theta}{2}e^{i\phi}, \sin\frac{\theta}{2}\right)$ corresponds to $U_S = S_b^2 - (0,0,+1)$. Equations (32) can also be written in formal mathematics:

$$s_N (r \in U_N \text{ on } S^2) = \left(\cos\frac{\theta}{2}, \sin\frac{\theta}{2}e^{-i\phi}\right); \quad s_S (r \in U_S \text{ on } S^2) = \left(\cos\frac{\theta}{2}e^{i\phi}, \sin\frac{\theta}{2}\right).$$

(33)

We have established the two cross sections of a particular point of $(z_1, z_2) \in S_R^3$, each corresponding to a chart on the $S_b^2$. We will now work on the inclusion map ($\beth°$) of these points on $S_R^3$. We have

$$\beth°\left(\cos\frac{\theta}{2}e^{i\varepsilon_1}, \sin\frac{\theta}{2}e^{i\varepsilon_2}\right) = \left(\cos\frac{\theta}{2}\cos\varepsilon_1, \sin\frac{\theta}{2}\cos\varepsilon_2, \cos\frac{\theta}{2}\sin\varepsilon_1, \sin\frac{\theta}{2}\sin\varepsilon_2\right)$$

(34)

$$\beth° s_N (r \in U_N \text{ on } S^2) = \beth°\left(\cos\frac{\theta}{2}, \sin\frac{\theta}{2}e^{-i\phi}\right) = \left(\cos\frac{\theta}{2}, 0, \sin\frac{\theta}{2}\cos\phi, -\sin\frac{\theta}{2}\sin\phi\right)$$

(35)

$$\beth° s_S (r \in U_S \text{ on } S^2) = \beth°\left(\cos\frac{\theta}{2}e^{i\phi}, \sin\frac{\theta}{2}\right) = \left(\cos\frac{\theta}{2}\cos\phi, \cos\frac{\theta}{2}\sin\phi, \sin\frac{\theta}{2}, 0\right)$$

(36)

The above simply means that the $\omega$ defined in terms of $(x_1, x_2, y_1, y_2)$ can be re-expressed in terms of $(\theta, \phi)$ on a re-parametrized $S_R^3$. In summary, the gauge corresponding to the North and South charts can be derived as follows:

$$A^{\mathcal{S}} = \frac{\hbar}{2e}(\beth°s_N) * \omega = \frac{\hbar}{2e}\left(-\cos\frac{\theta}{2}\sin\emptyset . d\left(\cos\frac{\theta}{2}\cos\emptyset\right) + \cos\frac{\theta}{2}\cos\emptyset . d\left(\cos\frac{\theta}{2}\sin\emptyset\right) - 0.d\left(\sin\frac{\theta}{2}\right) + \sin\frac{\theta}{2}.d(0)\right)$$

$$= -\frac{\hbar}{2e}(1+\cos\theta)d\emptyset$$

(37)

$$A^{\mathcal{N}} = \frac{\hbar}{2e}(\beth°s_N) * \omega = \frac{\hbar}{2e}\left(-0.d\left(\cos\frac{\theta}{2}\right) + \cos\frac{\theta}{2}.d(0) + \sin\frac{\theta}{2}\sin\emptyset . d\left(\sin\frac{\theta}{2}\cos\emptyset\right) + \sin\frac{\theta}{2}\cos\emptyset . d\left(-\sin\frac{\theta}{2}\sin\emptyset\right)\right)$$

$$= \frac{\hbar}{2e}(1-\cos\theta)d\emptyset$$

(38)

Noting the above, the gauge potential in the overlap regions are: $A^{\mathcal{N}} = A^{\mathcal{S}} + d\left(2\frac{\hbar}{2e}\emptyset\right)$. The transition functions are:

$$g_{SN} = \frac{z_2/|z_2|}{z_1/|z_1|} = e^{-i\emptyset} \; ; \qquad g_{NS} = \frac{z_1/|z_1|}{z_2/|z_2|} = e^{i\emptyset} .$$

(39)

The topological magnetic charge is $\Omega = \int dA^{\mathcal{N}} = \int dA^{\mathcal{S}} + dd\left(2\frac{\hbar}{2e}\emptyset\right) = \int dA^{\mathcal{S}}$. It is thus apparent that one could avoid the singular part of $A^{\mathcal{N}}$ by switching to $A^{\mathcal{S}}$ in the overlapping region, thus avoiding the conceptual difficulty of the Dirac string.